\documentclass[sigplan,10pt,final,screen,nonacm]{acmart}
\setcopyright{none}

\usepackage{amsmath}
\usepackage{amsfonts}
\usepackage{graphicx}
\usepackage{booktabs}
\usepackage{color}
\usepackage{algpseudocode}
\usepackage{listings}
\usepackage{multirow}
\usepackage{pgfplots}
\usepackage{pgf-pie}
\usepackage{url}
\usepackage{subcaption}
\usepackage{tikz}

\usepackage{algorithm}

\makeatletter
\g@addto@macro{\UrlBreaks}{\UrlOrds}
\let\OldStatex\Statex
\renewcommand{\Statex}[1][3]{%
  \setlength\@tempdima{\algorithmicindent}%
  \OldStatex\hskip\dimexpr#1\@tempdima\relax}
\makeatother

\pgfplotsset{compat=1.15}
\begin{document}

\title{MLGO: a Machine Learning Guided Compiler Optimizations Framework}

\author{Mircea Trofin}\authornote{These authors contributed equally.}
\affiliation{\institution{Google, Inc.}}
\email{mtrofin@google.com}

\author{Yundi Qian}\authornotemark[1]
\affiliation{\institution{Google, Inc.}}
\email{yundi@google.com}

\author{Eugene Brevdo}
\affiliation{\institution{Google, Inc.}}
\email{ebrevdo@google.com}

\author{Zinan Lin}
\affiliation{\institution{Carnegie Mellon University}}
\email{zinanl@andrew.cmu.edu}

\author{Krzysztof Choromanski}
\affiliation{\institution{Google, Inc.}}
\email{kchoro@google.com}

\author{David Li}
\affiliation{\institution{Google, Inc.}}
\email{davidxl@google.com}

\begin{abstract}
Leveraging machine-learning (ML) techniques for compiler optimizations has been widely studied and explored in academia. However, the adoption of ML in general-purpose, industry strength compilers has yet to happen.

We propose MLGO\footnote{We welcome your feedback! Please open an issue at \url{https://github.com/google/ml-compiler-opt} with the label \textit{paper}.}, a framework for integrating ML techniques systematically in an industrial compiler --- LLVM. As a case study, we present the details and results of replacing the heuristics-based inlining-for-size optimization in LLVM with machine learned models. To the best of our knowledge, this work is the first full integration of ML in a complex compiler pass in a real-world setting. It is available in the main LLVM repository. 

We use two different ML algorithms: Policy Gradient and Evolution Strategies, to train the inlining-for-size model, and achieve up to 7\% size reduction, when compared to state of the art LLVM -Oz. The same model, trained on one corpus, generalizes well to a diversity of real-world targets, as well as to the same set of targets after months of active development. This property of the trained models is beneficial to deploy ML techniques in real-world settings.

\end{abstract}

\maketitle

\section{Introduction}
\label{sec:intro}
Previous work \cite{vectorizerCGO2020, simon:2013} has shown promise in replacing compiler optimization heuristics with machine-learned policies. Heuristics are algorithms that, empirically, produce reasonably optimal results for hard problems, within pragmatic constraints (e.g. "reasonably fast"). In the compiler case, heuristics are widely used in optimization passes, even those leveraging profile feedback, such as inlining and register allocation. Such passes have significant impact on the performance of a broad variety of programs. These problems are often NP-hard and searching for optimal solutions may require exponential time or memory. Reinforcement Learning (RL) is a family of machine learning techniques that may be applied to find increasingly optimal solutions through an automated iterative exploration and training process.

Our focus is ahead-of-time (AOT) compilers, specifically, C/C++. In a real-world setting, we expect two main benefits from machine learning techniques: first, heuristics are human-trained based on a human-manageable set of benchmarks and regression cases. Machine learning easily scales to large corpora of training examples - which we expect to increase the likelihood of obtaining policies that generalize well. This is important because, as we will explore in detail, we do not want to retrain policies too frequently (it is an adoption blocker), nor do we want to train 'online', while the compiler is running in production (it would affect determinism). Second, heuristics are human-written code that needs to be maintained. This places a downward pressure on the number of program properties ("features") and the combinations between them that can be practically leveraged. We believe using more features and feature combinations would result in better optimization decisions. ML scales well with the addition of features, and can discover profitable feature combinations. While ML techniques may be able to address these two points, a trade-off is that maintaining and evolving them requires practices and approaches different from those used for heuristics.

As pointed out, applying ML to compiler optimizations has been explored by academia, but it has not been adopted in production environments. To explore why, we chose a pilot optimization problem and approached it with the intention to deploy in production. The goal of the pilot is to inform problem framing and design choices. Other than performing better than the tip-of-tree production compiler, we did not aim to advance the state of the art for the pilot problem.

The chosen problem is inlining-for-size in LLVM, and in particular, the inlining decision heuristics. The expectation was that this would offer representative challenges - size optimization is important for real-world scenarios, such as mobile software, and inlining is a particularly challenging optimization (see Section \ref{sec:llvm-inliner-overview}).

We chose size rather than speed for the pilot because size is relatively easy to measure and non-noisy, which we expected to aid in rapid prototyping by removing one source of potential problems (noisy rewards). We acknowledge that translating our experience to performance problems may seem non-immediate at a first glance, but we believe that not to be the case: for example, inlining for speed can be understood partially as a "for-size" problem, with respect to the instruction working set (more in Section \ref{sec:next-steps}).

\begin{figure}
    \centering
    \includegraphics[scale=0.4]{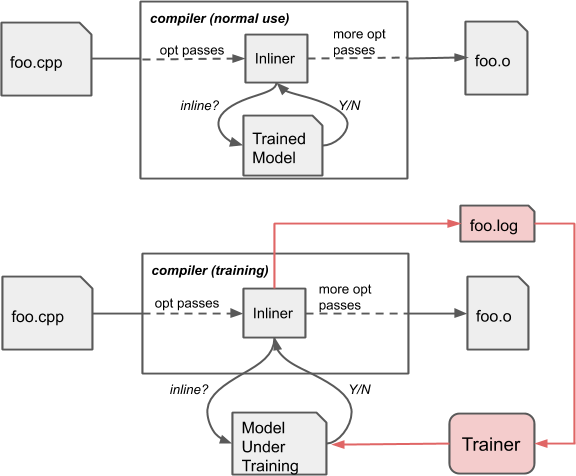}
    \caption{MLGO Overview}
    \label{fig:dev_overview}
\end{figure}

From a very high level, our MLGO framework separates the \textit{use} of the compiler from the \textit{training} of policies as shown in Figure \ref{fig:dev_overview}. Day-to-day production use is unchanged; as an implementation detail, a trained model embedded in the compiler is used to make decisions (in this case, inlining) that were previously handled by a manual heuristic. Training happens separately using a large, representative corpus of intermediate representation (IR) modules. The training process is iterative, each step using an updated policy, so the policy is not embedded in the compiler. During training, the inliner produces a log that records the inlining process (features, decisions, etc). The logs are collected and fed to the training algorithm to produce a new model.

The paper is organized as follows: Section \ref{sec:rl-overview} provides an overview of the relevant ML techniques and the LLVM inliner. Section \ref{sec:arch} gives an overview of MLGO: our framing of the problem of applying ML to compiler optimizations, and our methodology. Section \ref{sec:mldetails} describes the policy training in MLGO for the inlining problem. Section \ref{sec:llvm-impl} details the implementation, in LLVM, of our pilot project. The results are presented in section \ref{sec:results} and the related work is described in Section \ref{sec:related}. Finally in Section \ref{sec:next-steps}, we discuss our plans for applying the lessons learned so far to speed problems, as well as next steps in ML techniques that we are considering. Section \ref{sec:conclusion} concludes the paper.

A note: throughout this paper, we use interchangeably the terms \textbf{policy} and \textbf{model}. A compiler optimization policy is a decision rule that takes actions inside the optimization pipeline (e.g.,"should we inline this call graph edge or not?"). "Model" refers to a neural network implementing an optimization policy. Also, we use the term \textbf{heuristic} to refer to manually-crafted decision rules.

\section{Background}
\label{sec:rl-overview}
\subsection{Machine Learning Techniques for Replacing Compiler Optimization Heuristics}

There are two characteristics that make reinforcement learning (RL) a suitable tool for replacing compiler optimization heuristics: 1) there are no examples showing optimal strategies for these heuristics --- in the inlining problem, we don't know whether inlining or not for a certain call site is the optimal choice; 2) we can efficiently explore different strategies, and improve strategies from those experiences. The absence of examples ("labels") means we cannot use supervised learning. In contrast, RL is an area of machine learning that learns from trial and error instead of given labels. It has proven its success in robotics, playing Atari games, playing the game of Go, etc \cite{rl_atari, rl_robotics, rl_go}. In RL, an agent (i.e., the compiler) learns by repeatedly interacting with the environment (i.e., compiling) and gradually improves its policy (i.e., decision rules). More specifically, by compiling software again and again with different strategies, the compiler will come up with better and better policy on its own with RL algorithms. 

Previous work has shown Evolution Strategies (ES) to be a competitive alternative to RL algorithms in MuJoCo and Atari tasks \cite{es, es2}. Motivated by this, we also tried this method in compiler optimization problems. ES are a class of black box optimization techniques. Like RL, ES training is also able to gradually improve the strategy with trial and error, and thus is also a suitable tool for compiler optimization problems.

\subsection{High-level Overview of the Current LLVM Inlining Pass}
\label{sec:llvm-inliner-overview}

Today's LLVM inliner is a pass operating on a strongly-connected component (SCC) of the static call graph in a module\footnote{in LTO or ThinLTO mode, a module can consist of IR from multiple source files}, at a time in bottom up order. The inlined callee's call sites are added to a work list and iteratively considered for inlining in a top down fashion. A pipeline of optimizations (Inst combine, scalar replacement of aggregates (SROA), loop optimizations, etc) is then applied on each function in the SCC\footnote{The DAG walk is repeated up to a set of times if de-virtualization happens in cleanups}, after the SCC was processed. The effects of these optimizations impact inlining decisions of call sites in the SCC calling into the last one.

LLVM inlining consists of many heuristics: the choice of call site traversal, the set of "cleanup" function passes run on functions after they are modified because call sites were inlined, the timing of these cleanups, and finally, the decision to inline or not a specific call site.

The decision to inline or not a call site is itself built on top of a rich set of heuristics. The compiler first computes the static "cost" of the callee post inlining by traversing the callee body simulating post-inline cleanup passes. If some call site arguments are known to be constant at compile time, that information is used to evaluate what instructions / basic blocks would be simplified, should the inlining be carried out.  The computed cost is then compared with a threshold. The threshold value is based on things like call site hotness, inline keyword, etc. Bonuses are also given to callees with a single basic block or high percentage of SIMD instructions. In certain cases, the compiler may also choose to defer inlining if inlining the caller itself to its own callers first may result in better savings - it may be better to make a local non-optimal decision that, later, would open the opportunity for better optimizations due to more context being available, such as call parameters propagating from callers further up in the call graph.

These sets of heuristics have been tuned for years and our pilot project replaces the manual decision process described above with ML models.
\section{MLGO}
\label{sec:arch}

MLGO is a set of guidelines and requirements derived from our understanding of the problem of leveraging ML techniques for replacing manual optimization heuristics. We start with our understanding of the participating personas and their scenarios, which then motivates MLGO guidelines and design decisions.

\subsection{Persona: Compiler User}
This user wants to benefit from improved compiler optimizations. They care about: correctness and performance of the generated code; compilation determinism (i.e. identical output for identical input) - to leverage incremental builds; avoiding the added cost and complexity of new infrastructure requirements on build and release pipelines, such as new compiler run-time dependencies or new steps (like training); and timeliness of the build, as it impacts hardware resource planning and developer productivity\footnote{There are non-ML driven optimization alternatives that trade off significantly increasing compilation time for improved optimizations. An ML alternative needs to be competitive to justify its other trade-offs}. \textbf{Our goal is to introduce no changes to this user}.

To achieve this, the MLGO guidelines are:
\begin{enumerate}
\item To maintain correctness guarantees, we replace heuristics, not semantics-preserving code. For example, we change the decision making process for carrying out function inlining, not how the  inlining action is implemented. This is along the insight of separation of correctness vs policy observed earlier in \cite{stephensonPHD}
\item 'Online' training - meaning, training while the compiler is executing in production - is an anti-goal for us: it would hurt determinism and compilation performance. Instead, policy training happens offline. Trained policies are embedded in the compiler as statically linked native code, and the resulting compiler is subjected to the same release process it currently is. Build and release infrastructures and pipelines of targets using the compiler do not need to be changed. Build determinism using the ML-enabled compiler is ensured because the policies are fixed - no training happens when the compiler runs, only inference. While native compilation doesn't guarantee timeliness, it eliminates one source of concern. Due engineering diligence still needs to be applied  to ensure timely feature extraction, for instance.
\item We require ML techniques that yield policies that generalize well over different code bases and code changes, and do not need frequent retraining. The compiler user doesn't \textit{have} to worry about policy training (although they are free to do so and potentially get better results). This is akin to how, in the context of manual heuristics, a compiler user doesn't have to fine-tune passes (or author code in them) to get reasonable results. In particular, we do not see automated tuning of existing heuristic parameters as a viable solution. Tuning parameters have been available in compilers for a long time, and the experience has been that a set of values does not translate well from target to target. The policy, while adjustable, is still dominated by combinations/evaluations identified manually. In addition, requiring re-tuning would complicate product build and release pipelines, which we want to avoid on behalf of our user.
\end{enumerate}

We refer to this use of policies as \textbf{release mode} (since it is encountered by users of a released/shipped compiler).

\subsection{Persona: Compiler Engineer}
This user wants to drive better optimizations in the compiler, diagnose regressions, and incorporate findings: 
\begin{enumerate}
\item \textbf{Policy Creation.} The engineer wants to incorporate ML techniques in a compiler optimization pass.
\item \textbf{Policy Improvement.} Here, they investigate a specific regression encountered in production, or want to improve a ML-enabled pass.
\item \textbf{"The Ship Blocker".} The engineer must quickly resolve a ship-blocking regression introduced by a hot patch, and caused by a misbehaving RL-enabled policy.
\end{enumerate}

In all of these cases, the \textit{compiler engineer} improves a policy through repeated exploration and training (see Section \ref{sec:training-infrastructure}). They want flexibility in replacing the model under training, and have less concern with timeliness and determinism, especially since models under training may use small random perturbations, to facilitate exploration. We refer to this use of policies as \textbf{development mode}. Here, models are loaded via a command line option, the compiler may have extra runtime dependencies, and model evaluation may involve changes to the runtime behavior of the compiler --- because, for example, the model evaluators may be multi-threaded and/or JIT-ing.

Because of the tension between heuristic code complexity and hypothesized ability to improve the heuristic by incorporating more features (as discussed in the introduction), MLGO forgoes goals of human comprehensibility of the resulting policy (in contrast to \cite{simon:2013}). Instead, we focus on developing and evaluating alternative methodologies to address the above scenarios. We discuss our current understanding of the trade-offs, and expect that more clarity will arise as we apply the approach through the lifetime of a number of diverse projects. Section \ref{sec:mldetails} will detail our experience with developing \textit{Policy Creation}. We have less experience, at this point, with \textit{Policy Improvement} and \textit{"The Ship Blocker"}, and derive our direction from experiences in other domains.

\textit{Policy Improvement} is currently (i.e. for manual heuristics) an iterative engineering process. The trigger is typically regressions identified in the field. The compiler engineer diagnoses the problem, hypothesizes a solution, then ensures that the solution does not introduce regressions in some corpus of benchmarks; if regressions happen, the process is repeated. In the MLGO methodology, we envision a gradual process. We do not believe it presents significant negative trade-offs compared to the state of the art:
\begin{enumerate}
    \item Start by incorporating regression use-case(s) into training corpus and retrain the policy.
    \item If that fails, hypothesize missing features. This requires some manual diagnosing of the current policy behavior. While we treat the policy as a black box, we do observe its effects, and can formulate hypotheses as to what information may be missing - since the information we provide (features) is also observable. The needed skill set is close to what compiler engineers currently employ for manual heuristic development, and, just like for manual heuristics, evolving the feature set is likely an iterative process. Typically, adding features and retraining shouldn't result in regressions for the previous training corpus, which is a benefit of our approach over the manual heuristic case. We should note that, if applying feature auto-extraction \cite{cummins:pact2017, vectorizerCGO2020} proves feasible in production, this step collapses to the previous step.
    \item If the above also fails, involve an ML expert to investigate alternative training algorithms. This is akin to today's (rare) full pass rewrites (for example: new register allocation pass). The difference is the need for cross-disciplinary interaction. Our hope is that, with time and experience, MLGO offer a reusable library of best practices and training solutions available "off the shelf" to compiler engineers.
\end{enumerate}
\textit{Ship blockers} are those cases where the compiler engineer doesn't have the luxury to do deep investigations into compiler behavior, since they are on a tight time budget. Assuming the pathological case is identified (i.e. which compilation unit causes the compiler to misbehave), in the case of manual heuristics, the levers of control are: trial-and-error with different compilation flag values (change policy thresholds, for instance); modify user code (use inlining directives, for instance); or disable the specific optimization for a specific module.

In MLGO, the picture is similar. Other than policy thresholds, the control levers available to the engineer are the same. In addition, the engineer may choose to revert to an earlier version of the policy, or manual heuristics for the problematic module, and, if needed, experiment with threshold flags. Specific to ML-based policies, we are exploring with local training and overfitting: as we will detail, our experience so far shows that it is possible to train a policy on a single modern, multi-core workstation, and obtain a reasonably good result within a day. An engineer could attempt to specialize a policy to overfit for the pathological case, and compile that case with the specialized policy (while compiling the rest of the project with the non-overfitted policy). This is similar to "experimenting with flags", with the exception that the exploration is directed by a training algorithm and more likely to quickly converge to a solution. The trade-off is that changing heuristic flag values does not require a training infrastructure --- even if that infrastructure could be run locally.

\section{MLGO Policy Training}
\label{sec:mldetails}

In this section, we show how we use reinforcement learning (RL) and evolution strategies (ES) to train inlining policies in the MLGO framework. Sections \ref{sec:rl} and \ref{sec:es} present how we train the inlining policy for the inlining-for-size problem with RL and ES algorithms, and Section \ref{sec:pg-es} compares the pros and cons of the two algorithms. Section \ref{sec:training-infrastructure} concludes this section by giving an overview of our policy training infrastructure\footnote{\url{https://github.com/google/ml-compiler-opt}}. 

\subsection{Reinforcement Learning}
\label{sec:rl}
\subsubsection{RL Problem Formulation}
\label{sec:rl-formulation}
RL aims to find an optimal policy for a Markov Decision Process (MDP). MDP is a mathematical framework that models sequential decision making --- in inlining-for-size problem, we make sequential decisions whether to inline or not. An MDP can be represented by the tuple $<\mathcal{S}, \mathcal{A}, \mathcal{P}, \mathcal{R}>$ with state space $\mathcal{S}$, action space $\mathcal{A}$, state transition distribution $\mathcal{P}(s'|s, a)$, and reward function $\mathcal{R}(s, a)$. In the MDP formalism, at time $t$, the agent observes the state $s_t \in \mathcal{S}$ of the environment, then decides to take an action $a_t \in \mathcal{A}$.  It also receives the reward $r_t = \mathcal{R}(s_t, a_t)$.  The environment state then transitions to $s_{t+1} \in \mathcal{S}$ by sampling from the probability distribution $\mathcal{P}(s_{t+1}|s_t, a_t)$. This process repeats until the agent reaches a termination state at time $T$. The agent's decisions are a function (we call it policy) $\pi = Pr(a|s)$ that maps observed state $s$ to a distribution over actions. In our case, $\pi$ is a neural network and we call it policy network. RL algorithms aim to find the optimal policy $\pi^*$ to maximize the total reward\footnote{In general, the total discounted reward is $R = \sum_{t=0}^T \gamma^t r_t$, where $\gamma$ is the discounting factor; but we take $\gamma = 1$ so we ignore it here.} $R = \sum_{t=0}^T r_t$.

We first formulate the inlining-for-size problem as an MDP. The inlining pass traverses over the call sites in the call graph in a deterministic order and decides at each call site whether to inline or not. Every inlining operation changes the call graph. We treat this as a sequential decision process, and we formulate it into an MDP as:

\textbf{state $\mathcal{S}$}: we define the current call graph and the call site being visited to be the state.  

\textbf{action $\mathcal{A}$}: $\mathcal{A} = \{0, 1\}$, where $1$ means inline and $0$ means do not inline.

\textbf{state transition probability $\mathcal{P}$}: unlike usual MDPs, the state transition is deterministic (no randomness) in the inlining problem. After an action is taken (inline or not inline), the compiler determines what the next state is (updates the call graph and decides the next call site to visit).

\textbf{reward $\mathcal{R}$}: reward is defined to be the native size reduction after the action is taken. If $a = 0$ (do not inline), the reward is $0$ since nothing changes; if $a = 1$ (do inline), the reward is defined as:
\begin{equation}
S(Caller_{before}) - S(Caller_{after}) + 
\begin{cases}
S(Callee), &\text{callee deleted} \\
0, &\text{callee remains}
\end{cases}
\end{equation}

where $S(f)$ is the native size of function $f$. Note that we do not actually know what the native size would be for a certain function while performing inlining, since inlining operates at the IR level. The definition for reward here is not practical for training. We will discuss how we tackle this challenge next.

\subsubsection{Policy Gradient Algorithm}
Policy Gradient (PG) \cite{pg} is a family of RL algorithms derived
from REINFORCE \cite{reinforce}. Though we use Proximal Policy Optimization (PPO) \cite{ppo}, we first briefly introduce REINFORCE --- as PPO is an enhancement to REINFORCE and they work in very similar ways.

On a high level, all PG algorithms gradually improve the policy $\pi_\theta$ by computing the gradients of the parameters $\theta$ in the policy network w.r.t. the total reward $R$, and then update $\theta$ with the gradient to improve the policy. With $J(\theta)$ denoting the expected reward under policy $\pi_\theta$, the gradient $\nabla_{\theta} J(\theta)$ in REINFORCE is computed as:
\begin{equation}
\label{equation: pg_gradient}
\nabla_{\theta} J(\theta) = \mathbb{E} \left[ \sum_{t = 0}^{T} R \nabla_{\theta} \log \pi_{\theta}(a_t | s_t) \right]
\end{equation}

Here $\mathbb{E}$ is an expectation over the policy $\pi_\theta$ being applied to an inlining pass. In practice, this expectation is approximated with Monte Carlo methods --- with $n$ trajectories\footnote{A trajectory is defined as ($s_0, a_0, r_0, s_1, a_1, r_1, ... s_T, a_T, r_T$), and total reward $R=\sum_{t = 0}^T  r_{t}$} collected from compiling with policy $\pi_{\theta}$, the parameter $\theta$ is updated with:
\begin{equation}
\label{equation: pg}
\theta \leftarrow \theta + \alpha \frac{1}{n} \sum_{i = 1}^n \left\{ \sum_{t = 0}^T R_i \nabla_{\theta} \log \pi_{\theta} (a_{i, t}|s_{i, t}) \right\}
\end{equation}

where $\alpha$ is the learning rate. As $\theta$ is updated, the policy $\pi(\theta)$ tends to evolve in the direction that increases the total reward. Algorithm \ref{algorithm: pg} describes the process --- as training progresses, the policy gradually improves on its own by iterating between two stages: 1) compile with a new policy and collect fresh trajectories; 2) update policy network parameters $\theta$. 
\begin{algorithm}
\caption{MLGO PG Training Algorithm}
\label{algorithm: pg}
\begin{algorithmic}[1]

\State Initialize $\theta$

\For{iteration = 1, 2, ...}
  \State Compile with policy $\pi_{\theta}$ to collect $n$ trajectories 
  \State Update $\theta$ using Equation \ref{equation: pg}
\EndFor
    
\end{algorithmic}
\end{algorithm}

The details of training with PPO, which has several additional terms in the loss function, are available in \cite{ppo}. One core improvement of PPO is to subtract a baseline $B$ from the reward to reduce the variance. Equation \ref{equation: pg_gradient} is modified as:

\begin{equation}
\label{equation: ppo_gradient}
\nabla_{\theta} J(\theta) = \mathbb{E} \left[ \sum_{t = 0}^{T} (R - B) \nabla_{\theta} \log \pi_{\theta}(a_t | s_t) \right]
\end{equation}

Here the baseline $B$ describes what the $R$ is supposed to be (irrelevant to policy). By subtracting it, $R - B$ provides better information about the effectiveness of the policy $\pi_\theta$. 

The total reward $R$ in these equations can be replaced with the returns following action $a_t$: $\sum_{t' = t}^T r_{t'}$. In this case, the baseline $B$ is a value network $V(s_t)$ predicting the future returns $\sum_{t'=t}^T r_{t'}$ from the state $s_t$. We choose to use the total reward $R$ since: 1) it is directly available in the inlining-for-size problem, while partial returns would have to be approximated; 2) it is difficult to build the value network $V(s_t)$ with the reduced state. We will discuss the details in the next section.  

\subsubsection{Challenges and Implementation Details}
\label{sec:impletation details}
We run into two challenges when applying PPO to the inlining-for-size problem: 1) complex state space; 2) impractical reward definition.

\textbf{Complex state space: } Our MDP formulation defines the state as the current call graph and the call site being visited. Unfortunately, encoding and processing a call graph at each decision point may not be computationally practical for a general-purpose compiler to afford.

\textbf{Impractical reward definition:} It is difficult to know a function's native size $S(f)$ during the inlining pass because native code lowering happens in a later pass, and because its structure may change due to more of its call sites being inlined.

To tackle the first challenge, we approximate the true state by distilling the state space to $11$ numerical features as listed in Table \ref{table:feature-list}. These features describe the local call site and provide some global information about the call graph. Section \ref{sec:llvm-inl-size} details the features we use. We considered, but rejected for now, the use of (IR) code embedding techniques\cite{code2vec};  this allows us to minimize additional computational/memory costs. We plan to consider such techniques in the future. 

\begin{table}[h]
\begin{center}
\begin{tabular}{ |c|c|c| } 
\hline
Type & Feature \\
\hline \hline
\multirow{3}{4em}{caller feature} & caller\_basic\_block\_count \\ 
& caller\_conditionally\_executed\_blocks \\ 
& caller\_users \\ 
\hline
\multirow{3}{4em}{callee feature} & callee\_basic\_block\_count \\ 
& callee\_conditionally\_executed\_blocks \\ 
& callee\_users \\ 
\hline
\multirow{3}{4em}{call site feature} & callsite\_height \\ 
& cost\_estimate \\ 
& number\_constant\_params \\ 
\hline
\multirow{2}{4em}{call graph feature} & edge\_count \\ 
& node\_count \\ 
\hline
\end{tabular}
\end{center}
\caption{Features for Inlining for Size}
\label{table:feature-list}
\end{table}

One drawback of the simplified state is that it greatly reduces information available to the policy --- it only contains a part of the local call site information, and limited global call graph information. We do not expect this to hurt the policy network because it is roughly the same information available to the current inlining heuristic. However, this reduced state vector does not allow us to build the value network $V(s_t)$ baseline --- at a certain time $t$, the simplified state $s_t$ is not informative enough to predict the future return $\sum_{t'=t}^T r_{t'}$. 

A simple approach to side-stepping the lack of partial reward information and the side effect of reduced state representation is to use the total reward $R$ instead of the partial return as shown in Equation \ref{equation: pg_gradient}. While the per-step reward is difficult to get, the total reward is the sum of the individual (unknown) rewards --- it is relatively easy to evaluate: evaluate native size with / without inlining, and subtract. In the total reward setup, the baseline $B$ is defined as the estimated native size reduction of the module after the inlining pass. We can use the native size reduction under the heuristic policy as the baseline $B$.

Using the total reward instead of partial rewards has its drawbacks: 1) more data needs to be collected to achieve the same performance; 2) the final model quality may be worse.

\subsubsection{Warmstart with Behavioral Cloning Policy}
Instead of having the RL algorithm learn from the scratch (initialize $\theta$ randomly), we facilitate training by initializing $\theta$ from some "warmstart" policy. To facilitate the RL training, we need to have a "warmstart" policy that already performs reasonably well. An intuitive idea is the heuristic inlining decisions in LLVM. Therefore, we train the warmstart policy to imitate the heuristic inlining decisions in LLVM using behavioral cloning algorithm\cite{bc}. The behavioral cloning algorithm essentially views the problem as a supervised learning problem where the features are the same as the RL training and the label is the heuristic inlining decision --- it trains a neural network that makes inlining decisions as close as the heuristic inliner does. In this way, we get a policy that makes decisions similar to LLVM's current inlining heuristics and thus can serve as the warmstart policy to make our RL training much faster. 

\subsection{Evolution Strategies}
\label{sec:es}
Previous work has shown that ES, as a gradient-free black box optimization technique, is a competitive alternative to RL algorithms on MuJoCo and Atari tasks \cite{es}. 

ES focuses on black-box optimization problems of the form $\max_{\theta} F(\theta)$, where $F$ can be any \textit{black-box function} that can be evaluated. Given $\theta$, we essentially assume we have an oracle that calculates $F(\theta)$. In our specific case, $\theta$ are the parameters of the policy network $\pi_\theta$ and $F(\theta)$ is the total reward $R$ --- the native size reduction after inlining under policy $\pi_\theta$ for a certain module.

Instead of directly optimizing $F(\theta)$, ES focuses on optimizing $J(\theta)$, which is a smoothed version of $F(\theta)$:
\begin{equation}
\max_{\theta} J(\theta) = \max_{\theta} \mathbb{E}_{\varepsilon \sim \mathcal{N}(0, I)} F(\theta + \sigma \varepsilon).
\end{equation}

Here $\mathcal{N}(0, I)$ denotes the multivariate normal distribution with zero mean and identity covariance matrix. Similar to PG, ES also takes the gradient of the parameter $\theta$ w.r.t. $J(\theta)$:
\begin{equation}
\label{eq:es-gradient}
\nabla_{\theta} J(\theta) = \frac{1}{\sigma} \mathbb{E}_{\varepsilon \sim \mathcal{N}(0, I)} \{ F(\theta + \sigma \varepsilon) \varepsilon \} \;,
\end{equation}
and uses Monte Carlo approximation of the gradient to update $\theta$ to improve the policy:
\begin{equation}
\label{equation: es}
\theta \leftarrow \theta + \alpha \frac{1}{n \sigma} \sum_{i=1}^n \left\{ F(\theta + \sigma \varepsilon_i) \varepsilon_i \right\}
\end{equation}
where $\alpha$ is the learning rate, and $\varepsilon_i$ are vectors sampled from $\mathcal{N}(0, I)$.

Algorithm \ref{algorithm: es} describes the ES algorithm. Similar to PG, ES also iterates between data collection and policy update to gradually improve the policy.
\begin{algorithm}
\caption{MLGO ES Training Algorithm}
\label{algorithm: es}
\begin{algorithmic}[1]

\State initialize $\theta$

\For{iteration = 1, 2, ...}
  \State Sample $\varepsilon_1, \varepsilon_2, ..., \varepsilon_n \sim \mathcal{N}(0, I)$
  \State Compile with policy $\pi_{\theta + \sigma \varepsilon_i}$ to get $F(\theta + \sigma \varepsilon_i)$
  \State Update $\theta$ based on Equation \ref{equation: es}
\EndFor
    
\end{algorithmic}
\end{algorithm}

\subsection{PG v.s. ES}
\label{sec:pg-es}
While the policy gradient algorithm and the evolution strategies algorithm are similar on a high level, they are different in many ways and have their pros/cons.

\textbf{Complexity:} The key advantage of ES is that it is conceptually simpler: 1) it requires less engineering complexity --- unlike the PG algorithm where we need the logged trajectory during inlining $(s_1, a_1, s_2, a_2, ..., s_T, a_T)$ and the total reward $R$ for training, the ES algorithm only needs the total reward. Thus we do not need to log the trajectory while doing compilation; this reduces both engineering complexity and storage/network requirements; 2) it has less requirements on the problem structure as long as there is an oracle telling the total reward $F(\theta)$ under the policy parameters $\theta$. As a result, it is easier to apply ES to other compiler optimization problems as PG requires formulating the optimization problem as an MDP.

\textbf{Sample Efficiency:} Sample efficiency quantifies the amount of data required for training. The key advantage of PG is that it has much higher sample efficiency than ES. In the inlining-for-size problem, we observed that even though PG is trained using total reward, over $20X$ computational resources are needed to train an ES policy of similar quality. In problems where partial reward information is available after every decision point, we expect the sample efficiency gap to be even larger.

\subsection{Training Infrastructure}
\label{sec:training-infrastructure}
\begin{figure}[h]
    \centering
    \includegraphics[scale=0.35]{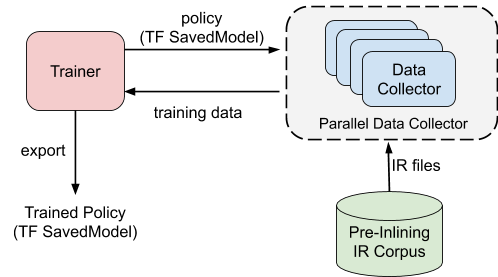}
    \caption{System Overview: Policy Training}
    \label{fig:training_infra}
\end{figure}

PG and ES algorithms are very similar in policy training on a high level --- both of them improve the policy $\pi_\theta$ by iterating between compiling with policy $\pi_\theta$ to collect data and update parameters $\theta$. Figure \ref{fig:training_infra} demonstrates their training workflow. Before the training, we prepare an IR corpus consisting of pre-inlining IR files extracted from some software. At each iteration, the trainer sends the policy $\pi_\theta$ to data collector, the data collector samples several IR files from the IR corpus, does compilation to collect training data, and sends the training data back to the trainer for training. The training is done after several iterations, and the trainer exports the trained policy. We use TF-Agents\cite{TFAgents} --- an RL library in TensorFlow\cite{tensorflow} for training and the policy is in the format of TensorFlow SavedModel.

The bottleneck of training for the inlining problem is data collection. Therefore, the data collection is carried out in a parallel way to improve the overall training efficiency.

\begin{figure}[h]
    \centering
    \includegraphics[scale=0.35]{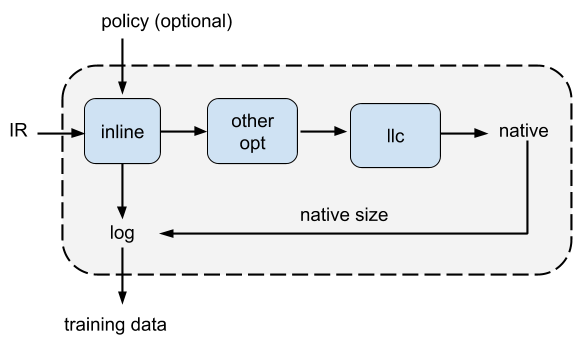}
    \caption{Data Collection for Inlining-for-Size}
    \label{fig:inline}
\end{figure}

Figure \ref{fig:inline} details how the data collector module works. It is supported by the development mode in MLGO framework. It takes a pre-inlining IR file and a policy (optional) as inputs, conducts inlining on the IR file based on the policy, has the post-inlining IR file optimized by other opt passes after inlining, converts the optimized IR file into native code, and gets the native size of this module. The native size, together with the log file generated during inlining by MLGO that contains the trajectory ($s_1, a_1, s_2, a_2, ..., s_T, a_T$), composes the output of the data collector module --- training data. If the policy is not given, the inliner will conduct the current heuristic inlining and log the trace. It has two use-cases as discussed in Section \ref{sec:rl}: 1) collect data to train the warmstart policy with behavioral cloning algorithm; 2) use the heuristic inlining as the baseline. The ES algorithm only needs the reward (native size) for training so the log file is not needed.

\section{LLVM Implementation}
\label{sec:llvm-impl}

We implemented the pilot project in LLVM\footnote{Code references are made in the context of commit 71059257bd4.}, together with reusable support for release and development modes, as well as continuous integration build bots.

\subsection {RL Driven Inlining (for Size) in LLVM}
\label{sec:llvm-inl-size}

We introduced an abstraction for the inline decision-making policy, the \textit{InlineAdvisor}, and a module analysis, \textit{InlineAdvisorAnalysis}, that may be used to retrieve the InlineAdvisor. The analysis can not be accidentally invalidated by other passes. This is necessary, since the inliner pass is interleaved with the execution of function passes, as previously discussed, and we want to track module-wide features throughout the performance of inlining and related passes over a module. Instead, the analysis is managed explicitly - see ModuleInlinerWrapperPass (llvm/Transforms/IPO/Inliner.h). The specific implementation of the advisor is chosen through a LLVM flag (-enable-ml-inliner). By default, the implementation is the manual heuristic. Passing 'release' or 'development' to the flag selects the respective mode, if the compiler was built with support for that mode.

Feature extraction is modeled as a separate analysis, \textit{FunctionPropertiesAnalysis}, and reused by the release and development implementations. The full feature set is captured in llvm/Analysis/InlineModelFeatureMaps.h. We capture some call site-local information, as well as global information, such as module-wide number of functions and static calls, caller/callee user counts; position in the original call graph, as the distance of the call site to the farthest SCC; and an estimate of removed instructions given the call site context.

We use TensorFlow \cite{tensorflow} as the model training and inference framework. In both modes, the interface between LLVM and the model is defined solely in terms of input and output tensor specifications (tensor name, type, and shape). The internal structure of the model is an implementation detail. This means that during training, the compiler engineer is free to explore hyper-parameters or add/modify hidden layers. Also, ingesting a new model with a different internal structure, in release mode, is just a matter of recompiling LLVM.

Refer to lib/Analysis/\{MLInlineAdvisor | ReleaseModeModelRunner | DevelopmentModeInlineAdvisor\}.cpp for more implementation details. At a high level, both release and development modes:
\begin{itemize}
\item Handle user inlining directives and correctness aspects (these are done without model evaluation)
\item Extract the features \textit{associated} with a call site and form fixed-sized tensors (primitive data type vectors), and efficiently maintain the module-wide features
\item Pass the tensors to the model evaluator, and request it perform an evaluation
\item Take the result of the evaluation as advice (i.e. inline/don't) and make that available to the inliner pass.
\item It is possible that a policy misbehaves in unforeseen circumstances (which, as a note, should then be incorporated into the training loop). The resulting IR, while correct, could become increasingly expensive to process by subsequent passes. To avoid this, we set a hard threshold to the amount by which the number of instructions may grow in a compilation unit.
\end{itemize}

\subsection{Release Mode Implementation in LLVM}
\label{sec:relmode}

The model is encoded in the TensorFlow serialization format, SavedModel \cite{savedmodel}, which is compiled into native code by the saved\_model\_cli tool \cite{tfcompile}. To use this tool, we added a build rule (see llvm/cmake/modules/TensorFlowCompile.cmake) to the LLVM build system. Applying the rule to a model generates a header file and an object file. From here, the model may be consumed as a C function; for simplicity, the SavedModel compiler provides a thin C++ wrapper, exposing plain C/C++ APIs (primitive types), which is compiled as part of the LLVM build process. The SavedModel is checked in as source\footnote{The SavedModel separates the evaluation graph structure from the value of the trained weights used for evaluation. The graph is stored as text. The weights/serialized float arrays are stored as a binary blob. Their evolution (due to training) does not diff well, so the compactness of a binary format is more economical for the project repository.}.

To build with support for the release mode, the SavedModel compiler must be available during the build time of LLVM. The compiler may be installed through a python pip package\footnote{See the buildbot setup script available at \url{https://github.com/google/ml-compiler-opt/blob/58bf347286c21519b3cc418f659c485cbb7ad82f/buildbot/buildbot_init.sh}}. Once installed, its location is provided to the LLVM build via the TENSORFLOW\_AOT\_PATH cmake flag. Specifying that flag also defines a conditional compilation flag, HAVE\_TF\_AOT, which enables the compilation as part of the Analysis component of the support for release mode.

Note that this mechanism would build the release mode implementations of \textit{all} optimization passes that have RL-driven policies, meaning that implementers just need to reuse the same mechanisms - conditional compilation flag, build rule, etc - to plug in a ML-based policy replacement.

\subsection{Development Mode Implementation in LLVM}
\label{sec:devmode}

As discussed, in development mode, we want to support loading models from the command line. For the development mode, the build time dependency is to the TensorFlow C API library, instead of the TensorFlow pip package. Model loading, initialization, and evaluation is performed via a reusable C++ API wrapper (see lib/Analysis/TFUtils.cpp) that simplifies the programming model of this aspect of development mode implementations.

In addition to facilitating a different model ingestion mechanism, the development mode is responsible for producing traces necessary for training ("training logs")\footnote{RL algorithms require these. ES algorithms do not}. These logs capture the succession of feature values observed when the policy is asked to make a decision, and the decision made afterwards ("trajectories"). Training logs may be produced for both the heuristic policy (for bootstrapping training - "warmstart") as well as for the ML policy currently under training. Exploration - i.e. deviating from policy, with the purpose of finding new learning opportunities - is delegated to a TensorFlow mechanism that introduces some randomness in decisions. This mechanism is an implementation detail of the model as produced by the training algorithm, and is outside the control of the compiler. Care must be taken to remove such randomness before shipping a model, and re-validate its effectiveness. We encode the training logs as textual SequenceExamples\cite{tf.SequenceExample} proto-buffers, the typical abstraction Tensorflow training algorithms would expect. We produce a textual output to avoid an additional dependency to LLVM, and to simplify diagnostics and testing of the feature.

Analogous to the release mode, enabling development mode in LLVM requires the dependency be made available to the build system. In this case we use the TENSORFLOW\_C\_API flag, which in turn defines the HAVE\_TF\_API conditional compilation flag. More details may be obtained from the previously noted build bot scripts. Also similar to the release mode, this mechanism enables all cases that have the TensorFlow C API library dependency. Unlike release mode, the development mode's use of the TensorFlow C library is a run-time dependency, and needs to be on the loader path.
\section{Evaluation}
\label{sec:results}
\subsection{Compilation Overhead}
Model evaluation in release mode has \textbf{fixed cost}, both in terms of compiler run-time memory utilization, as well as CPU utilization. This is because models are fixed size graphs connecting functional operators, taking fixed sized inputs, using constant weights, and producing fixed sized outputs. For the current model, we observed ~0.65\% increase in memory utilization at run-time. When inlining a large IR module (~33MB), we measured a 10\% increase in \textbf{inlining} time, mostly attributable to feature extraction; since inlining tends to represent ~10-15\% of \textbf{total} compile time, the net contribution of the release mode is only 1\%. Finally, clang binary size increase due to the inclusion of the compiled model was ~115KB, representing 0.08\% size increase.

We did not formally measure the overhead of the development mode, mainly because timeliness is less of a concern here, and also because model evaluation may happen through a variety of means, including JIT-ing, which makes measurements more unstable. We did want to validate that the solution is practically usable in training loops, and observed ~26K IR modules being inlined in parallel on a 72 thread machine, 192GB RAM, in around 10 minutes, and without going past half of the available RAM\footnote{Anecdotally, we were able to built Fuscia using a development mode clang, and timeliness was not a noticeable issue.}.

\subsection{Inlining for Size Results}
We trained the inlining for size policy on an internal search application containing over $28000$ IR modules with a variety of different code patterns\footnote{We also have an end-to-end demo at  \url{https://github.com/google/ml-compiler-opt/blob/main/docs/demo/demo.md} that trains on publicly available code and achieves similar performance.}. The rich set of patterns will improve generalizability, across both time and software domain, of the trained policy. As mentioned, this is important for real-world deployment.

We trained the policy using both PG and ES on the internal search software. Table \ref{table:basic_performance} compares their effectiveness in terms of reduction of the .text section compared with heuristic-driven -Oz. We trained $3$ policies: PG and ES with a $2$ hidden layer $(40, 20)$ neural network, and ES(L) with a deeper $4$ layer $(20, 20, 20, 20)$ neural network\footnote{Detailed hyper-parameters at \url{https://github.com/google/ml-compiler-opt/tree/main/compiler_opt/rl/inlining/gin_configs}}. We can see that: 1) PG has better sample efficiency than ES --- it consumes \textasciitilde $5\%$ training resources of ES ($100 * 12$ v.s. $488 * 60$); 2) better policies may be achieved with a larger neural network at the cost of more training resources; 3) we can train a reasonable PG policy on a single multi-core (e.g. $72$ threads) high-performance machine well within a day.
\begin{table}
\begin{center}
\begin{tabular}{ |c|c|c|c| } 
\hline
 & PG & ES & ES (L) \\
\hline \hline
Size Reduction & $4.95\%$ & $3.74\%$ & $5.94\%$ \\
\hline
Parallelism in Data Collection & $100$ & $488$ & $488$ \\
\hline
Training Time & \textasciitilde $12$h & \textasciitilde $60$h & \textasciitilde $150$h \\
\hline
\end{tabular}
\end{center}
\caption{Policy Gradient v.s. Evolution Strategies}
\label{table:basic_performance}
\end{table}

\subsubsection{Generalizability across Software}
\begin{figure}[h]
    \centering
    \includegraphics[scale=0.3]{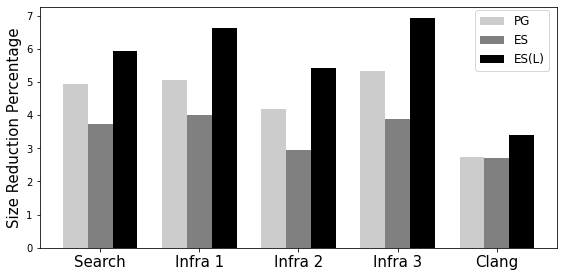}
    \caption{Generalizability across Software}
    \label{fig:internal}
\end{figure}

\begin{figure*}
    \centering
    \includegraphics[scale=0.3]{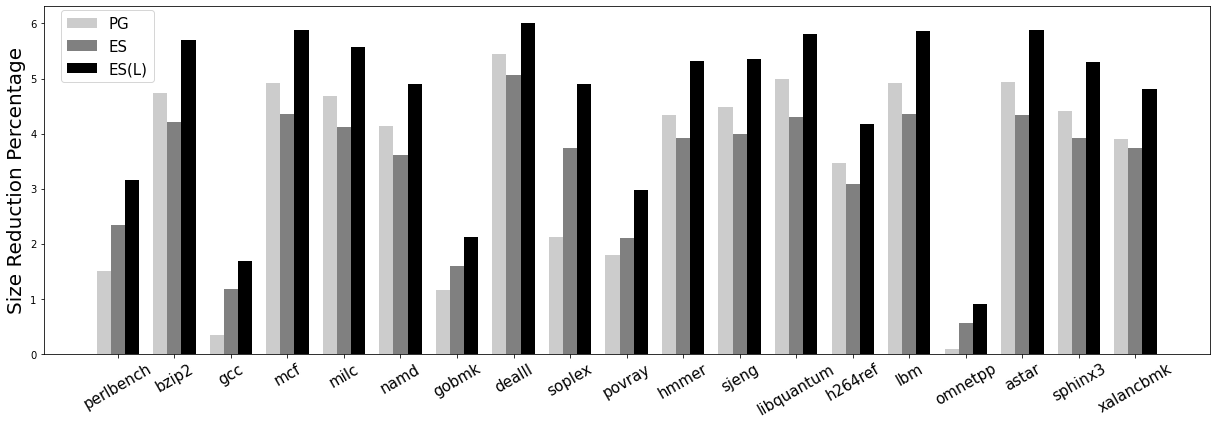}
    \caption{SPEC 2006 Size Reduction}
    \label{fig:spec}
\end{figure*}

We deploy the trained PG and ES policies to a wide range of software to evaluate their generalizability. Figure \ref{fig:internal} shows how the $3$ models we trained on the search application perform on $3$ different internal applications and on Clang\footnote{Specifically, clang @4ca60915bcc (2020/8/28) building clang @d469133f95b (2020/4/25).}. Figure \ref{fig:spec} shows their effectiveness on SPEC 2006. We can see that all the $3$ policies show good generalizability --- they are able to reduce the native size to some extent. Policy effectiveness is ES(L)$>$PG$>$ES for most software, which is the same as what we see on the search application. It also suggests good generalizability as a policy performs better on a certain software is likely to also perform better on other software.

\subsubsection{Generalizability across Time}
\begin{figure}[h]
    \centering
    \includegraphics[scale=0.3]{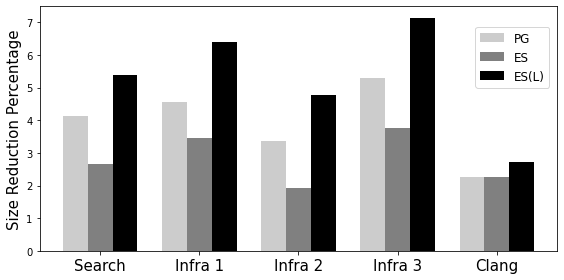}
    \caption{Generalizability across Time}
    \label{fig:time}
\end{figure}

To evaluate the trained policies' generalizability across time, we deploy the $3$ trained policies on the same software as in Figure \ref{fig:internal} $4$ months later. We also use the LLVM $4$ months later\footnote{Clang selfhost @4ca60915bcc (2020/8/28).}. Both the software and the compiler have been under active development in that period. Figure \ref{fig:time} demonstrates the results. We can see that their effectiveness may degrade somewhat (compared with Figure \ref{fig:internal}), but they still have decent wins compared with the current -Oz.

\section{Related Work}
\label{sec:related}

There have been many academic efforts in using machine learning and related techniques to replace hand-crafted heuristics in compilers. Our contribution is identifying the problem framing and design constraints that enable applying these techniques to production.

Wang and O'Boyle~\cite{wang:2018} present an extensive survey of the use of machine learning in compiler optimizations. Most, however, employ supervised learning techniques, which, as explained, are not in our scope. The closest, Cavazos et all - \cite{cavazos2005automatic} used unsupervised learning to automatically tune the inlining parameters (thresholds) of a research Java Virtual Machine (JVM), which features a very simple manually written heuristic. In subsequent work, Simon et al.~\cite{simon:2013} construct a heuristic as a decision tree, to address maintainability and evolvability. While the ML techniques are similar to what we are using in MLGO, both parameter tuning and direct policy comprehensibility are counter to our goals, as described in section \ref{sec:arch}.

Adams et al. \cite{halide:19} employed ML to train a cost model to automatically schedule Halide programs for image processing. With runtime sampling, the cost model is used to find the optimal schedule parameters using beam search. Similarly, Chen et al.\cite{autotvm:18} used deep learning to train a statistic model for TensorFlow programs.

Inlining-specific, Dean et al. \cite{Dean93trainingcompilers} build a database of observed decisions and their effects, and consult it in subsequent compiler runs - while this is not learning, it is a precursor of efforts in this area. In \cite{cooper:08}, Cooper et al. presented a scheme to parameterize the inline heuristics (decision tree) and the hill-climbing parameter space search. 

Haj-Ali et al. \cite{vectorizerCGO2020} use reinforcement learning to instrument source code with pragma directives to drive the vectorization pass. The policy does not replace a compiler heuristic, rather it informs one, by augmenting source code as a pre-build step. This does not make the technique transparently deployable for compiler users. That being said, we currently see no fundamental reason their solution cannot be adapted to MLGO. The main practical issue we see is understanding trade-offs of automated feature extraction, which we intend to explore as a next step as well.

Supervised learning is used by Stephenson and Amarasinghe~\cite{stephenson:2005}  to predict loop unrolling factors and by Eliot et al.~\cite{eliot:nips1997} to train a local (single basic block) instruction scheduler. It uses a machine model to predict the so called preference relationship given a partial schedule and two candidate/ready instructions.  Cummins et al.~\cite{cummins:pact2017} automatically extract features from source code, and use supervised learning to learn heuristics for  predicting optimal mapping for heterogeneous parallelism and GPU thread coarsening factors.

Instruction scheduling is a hard problem in the compiler that extensively uses heuristics. The application of learning to instruction scheduling within straight line code has been explored by Moss et al.~\cite{moss:1998} and McGovern et al.~\cite{mcgovern:2002}. 

Data prefetching plays a similar role in bringing \emph{data} into the processor without stalls. In ~\cite{milad:icml2018}, Hashemi et al. treated the memory prefetching strategies as an n-gram classification problem in natural language processing, and used LSTM based Recurrent Neural Network (RNN) to do the prediction. Peled et al.~\cite{peled:2015} define the notion of \emph{semantic locality} and use reinforcement learning techniques to build a context-based memory prefetcher that approximates semantic locality. 

Another approach to optimize programs without dealing with specific optimizations is \emph{super-optimization}. This refers to the process of finding a better version of a given program that is semantically equivalent. Early efforts in super-optimization relied on brute force search. Recent efforts have focused on using stochastic search to improve the efficiency. Bunel et al.~\cite{bunel:2016} have used reinforcement learning to optimize stochastic search based super-optimization techniques.

Milepost GCC \cite{milepost} is a self-tuning GCC-based compiler, where program features are used to predict compiler flags beneficial to some goal (such as performance or size). It does not use ML-trained policies as part of its implementation.
\section{Future Directions}
\label{sec:next-steps}
\subsection{Applying MLGO to Speed Optimizations}

The immediately-observable difference between our pilot project and speed problems is that the reward is measured differently: speed is measured through benchmark runs, which are more time consuming and more noisy than size measurements. Using benchmark runs results as reward for speed optimization will have difficulties scaling, so our current preference is to avoid benchmark runs altogether, and focus instead on using problem-specific reward approximations. 

For register allocation, for example, a natural reward is calculating, per function, the block frequency-weighed sum of introduced moves. For inlining for speed, we plan to use a linear combination of a per-critical call graph estimate of working set (i.e. cache lines needed for execution) and dynamic instruction count. Both approaches require profiling information for carrying out the analysis, which we assume as a pre-requisite for workloads that are concerned with speed.

\subsection{ML Techniques}
There are multiple directions to pursue in terms of the ML techniques:

\textbf{Richer State Representations:} instead of using the $11$ numerical features to represent the state, we can have richer state representations. For example, we can use code embedding techniques \cite{code2vec} to embed the caller/callee to get more detailed information about the call site; or we can use graph neural network techniques \cite{graphsage} on the neighboring sub-graph of the call site to get more information about the call graph.

\textbf{PG with Partial Reward}: PG with partial reward would greatly improve the sample efficiency and trainability. However, there are two challenges to tackle: 1) find an efficient way to encode the global call graph information into state; 2) train a supervised model to predict a function's native size from its IR. 
\section{Conclusion}
\label{sec:conclusion}

We investigated the problem of leveraging ML techniques for compiler optimization in a real-world setting. We proposed a particular understanding of the problem space, and derived the MLGO framework. We applied it to inlining-for-size and described the resulting implementation, available in LLVM as a build-time opt-in, as well as the training methodology, two training algorithms and their trade-offs, and results. We are currently applying the same principles to addressing inlining for speed and register allocation policies, and hope that, through our experience, as well as that of the community, we can further refine MLGO and eventually mature it to a solution that compiler engineers can broadly apply and leverage machine learning for compiler optimizations in real-world settings.

\bibliography{main}

\end{document}